\documentclass[12pt,letterpaper]{JHEP3}
\usepackage{cite}
\usepackage{epsfig}
\usepackage{amsmath}
\usepackage{amssymb}
\usepackage{subfigure}

\title{Eternal inflation predicts that time will end}

\author{Raphael Bousso$^{a,b,c}$, Ben Freivogel$^{d}$, Stefan Leichenauer$^{a,b}$ and Vladimir Rosenhaus$^{a,b}$\\ \\
  $^a$ Center for Theoretical Physics and Department of Physics\\
\ \  University of California, Berkeley, CA 94720-7300, U.S.A.\\
$^b$  Lawrence Berkeley National Laboratory, Berkeley, CA 94720-8162,
  U.S.A.\\
$^c$  Institute for the Physics and Mathematics of the Universe\\ 
\ \ University of Tokyo,
5-1-5 Kashiwa-no-Ha, Kashiwa City, Chiba 277-8568, Japan\\
$^d$\! Center for Theoretical Physics and Laboratory for Nuclear Science\\
\ \ Massachusetts Institute of Technology, Cambridge, MA 02139, U.S.A.}

\abstract{Present treatments of eternal inflation regulate infinities
  by imposing a geometric cutoff.  We point out that some matter
  systems reach the cutoff in finite time.  This implies a nonzero
  probability for a novel type of catastrophe.
  According to the most successful measure proposals, our galaxy is likely
  to encounter the cutoff within the next 5 billion years.}

\begin{document}

\section{Time will end}
\label{sec-intro}
A sufficiently large region of space with positive vacuum energy will expand at an exponential rate.  If the vacuum is stable, this expansion will be eternal.  If it is metastable, then the vacuum can decay by the nonperturbative formation of bubbles of lower vacuum energy.  Vacuum decay is exponentially suppressed, so for a large range of parameters the metastable vacuum gains volume due to expansion faster than it loses volume to decays~\cite{GutWei83}.  This is the simplest nontrivial example of eternal inflation.

If it does occur in Nature, eternal inflation has profound implications.  Any type of event that has nonzero probability will happen infinitely many times, usually in widely separated regions that remain forever outside of causal contact.  This undermines the basis for probabilistic predictions of local experiments.  If infinitely many observers throughout the universe win the lottery, on what grounds can one still claim that winning the lottery is unlikely?  To be sure, there are also infinitely many observers who do not win, but in what sense are there more of them?  In local experiments such as playing the lottery, we have clear rules for making predictions and testing theories.  But if the universe is eternally inflating, we no longer know {\em why\/} these rules work.  

To see that this is not merely a philosophical point, it helps to consider cosmological experiments, where the rules are less clear.  For example, one would like to predict or explain features of the CMB; or, in a theory with more than one vacuum, one might wish to predict the expected properties of the vacuum we find ourselves in, such as the Higgs mass.  This requires computing the relative number of observations of different values for the Higgs mass, or of the CMB sky.  There will be infinitely many instances of every possible observation, so what are the probabilities?  This is known as the ``measure problem'' of eternal inflation.

In order to yield well-defined probabilities, eternal inflation requires some kind of regulator.   Here we shall focus on geometric cutoffs, which discard all but a finite portion of the eternally inflating spacetime.   The relative probability of two types of events, 1 and 2, is then defined by 
\begin{equation} 
\frac{p_1}{p_2}=\frac{\langle N_1 \rangle}{\langle N_2 \rangle}
\label{eq-p1p2}
\end{equation}
where $\langle N_1 \rangle $ is the expected number of occurences of the first type of event within the surviving spacetime region.  (We will drop the expectation value brackets below for simplicity of notation.) Here, 1 and 2 might stand for winning or not winning the lottery; or they might stand for a red or blue power spectrum in the CMB.  The generalization to larger or continuous sets of mutually exclusive outcomes is trivial.

\begin{figure}[tbp]
\centering
\subfigure[Global Cutoff]{
   \includegraphics[width=2.925in]{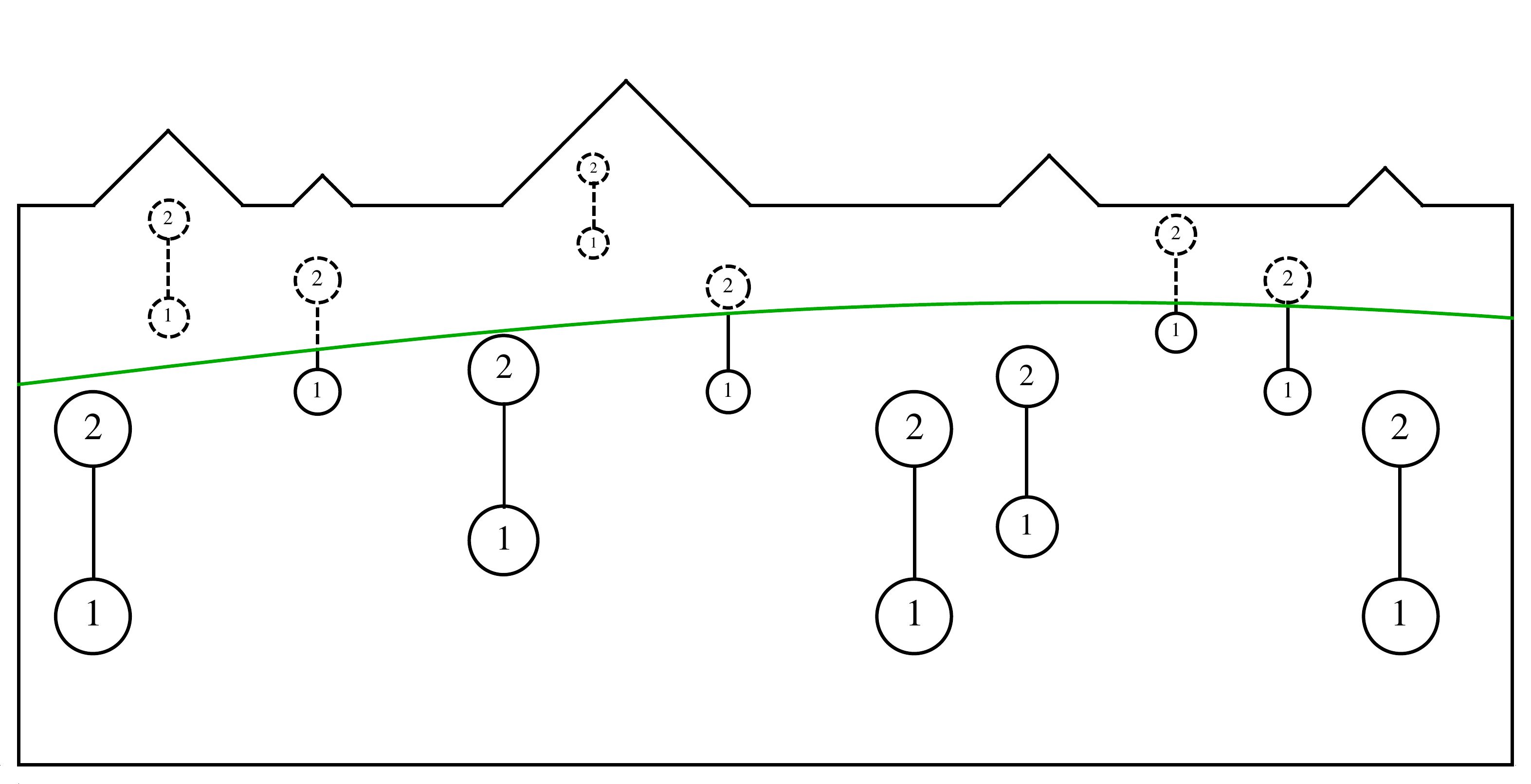}
   }
    \subfigure[Causal Patch Cutoff]{
    \includegraphics[width=2.925 in]{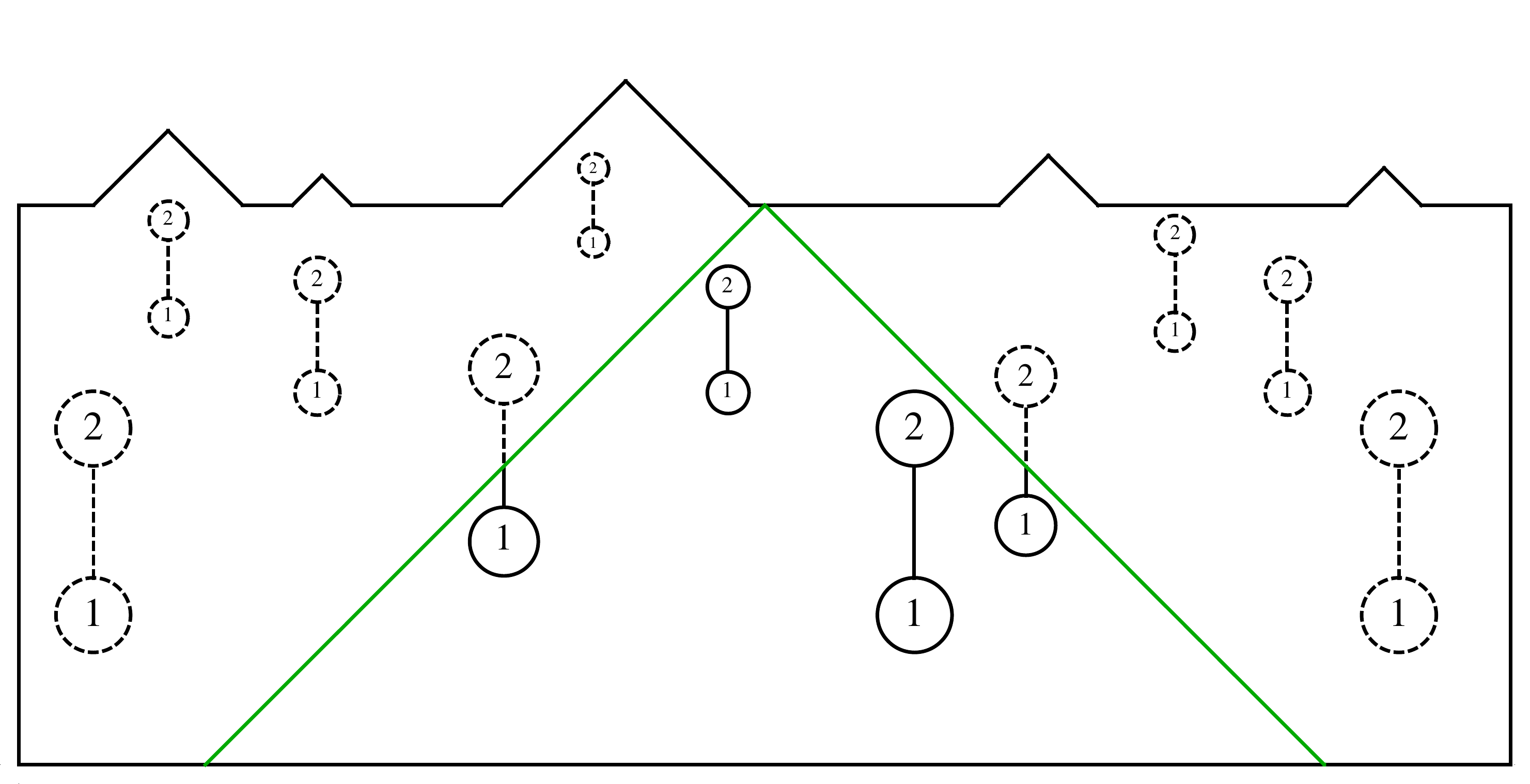}
    }
    \caption{A multiverse populated by infinitely many observers (vertical line segments) who first see 1 o'clock (at events labeled ``1'') and then 2 o'clock (``2'').  A geometric cutoff selects a finite set of events, whose relative frequency defines probabilities.  Events that are not counted are indicated by dashed lines.  The left figure shows a global cutoff: all events prior to the time $t_0$ (curved line) are counted and all later events ignored. (The global time has nothing to do with the observers' clocks, which come into being at a rate dictated by the dynamics of eternal inflation.) The right figure shows the causal patch cutoff, which restricts to the causal past of a point on the future boundary.  In both figures, the cutoff region contains observers who see 1 o'clock but not 2 o'clock.  Their number, as a fraction of all observers who see 1 o'clock before the cutoff, defines the probability of reaching the end of time between 1 and 2 o'clock.}
   \label{fig-end} 
\end{figure}

There are different proposals for what spacetime region should be retained.  Our basic observation in this paper applies to all geometric cutoffs we are aware of, and indeed seems to be an inevitable consequence of any simple geometric cutoff: Some observers will have their lives interrupted by the cutoff (Fig.~\ref{fig-end}).

Let events 1 and 2 be the observation of 1 o'clock and 2 o'clock on an observer's watch.  For simplicity, we will suppose that local physics can be approximated as deterministic, and we neglect the probability that the observer may die between 1 and 2 o'clock, or that the clock will break, etc.  Each observer is born just before his watch shows 1 and dies just after it shows 2, so that no observer can see more than one event of each type.

Conventionally, we would say that every observer sees both 1 o'clock
and then 2 o'clock.  But the figure shows that for some observers, 2
o'clock will not be under the cutoff even if 1 o'clock is.  A fraction
$1-N_2/N_1$ of observers are prevented from surviving to 2 o'clock.
The catastrophic event in question is evidently the cutoff itself: the
observer might run into the end of time.

One can imagine a situation where the relative number of observations
of 1 o'clock and 2 o'clock is relevant to predicting the results of an experiment. Suppose that the observers fall asleep right after seeing 1
o'clock.  They wake up just before 2 o'clock with complete memory
loss: they have no idea whether they have previously looked at their
watches before.  In this case, they may wish to make a prediction for
what time they will see.
 Since $N_2 < N_1$, by Eq.~(\ref{eq-p1p2}), $p_2<p_1$: the observation
 of 2 o'clock is less probable than that of 1 o'clock. This is
 possible only if some observers do not survive to 2 o'clock.

The conclusion that time can end obtains whether or not the
observers have memory loss. Consider an observer who retains her
memory.  She is aware that she is about to look at her watch for the
first time or for the second time, so the outcome won't be a surprise
on either occasion.  But this does not contradict the possibility that
some catastrophic event may happen between 1 and 2 o'clock.  The
figure shows plainly that this event {\em does\/} happen to a nonzero
fraction of observers.  The only thing that changes when considering
observers who remember is the type of question we are likely to ask.
Instead of asking about two alternative events (1 or 2), we may find
it more natural to ask about the relative probability of the two
different possible histories that observers can have.  One
history,``$1-$'' , consists of seeing 1 o'clock and running into the
cutoff.  The alternative, ``$12$'', consists of seeing 1 o'clock and
then seeing 2 o'clock.  From Fig.~\ref{fig-end} we see that
$N_{12}=N_2$ and $N_{1-}=N_1-N_2$.  Since $N_1>N_2$, we have
$p_{1-}>0$: there is a nonzero probability for the history in which
the observer is disrupted by the end of time.

\paragraph{Outline and Frequently Asked Questions}

The probability for the end of time is nonzero for all geometric cutoffs.  Its value, however, depends on the cutoff.  In Sec.~\ref{sec-rate} we compute the probability, per unit proper time, that we will encounter the end of time.

A number of objections may be raised against our conclusion that time can end.  \begin{itemize}
\item Q: Can't I condition on the end of time not happening?\footnote{In the above example, this would force us to ask a trivial question (``What is the relative probability of seeing 1 or 2, for an observer whose history includes both 1 and 2?''), which yields the desired answer ($p_2/p_1=1$).  For a more interesting example, consider an experiment that terminates at different times depending on the outcome, such as the Guth-Vanchurin paradox described in Sec.~\ref{sec-GV}, or the decay of a radioactive atom.  In such experiments it is tempting to propose that the experiment should be regarded to last for an amount of time corresponding to the latest possible (conventional) outcome, regardless of the actual outcome; and that any outcome (other than the end of time) should be counted only if the experiment has been entirely completed before the cutoff, in the above sense.  This proposal is not only futile (as described in the answer), but also ill-defined, since any event in the past  light-cone of the event $P$ can be regarded as the beginning of an experiment that includes $P$.}

A: Certainly.  This is like asking what the weather will be tomorrow, supposing that it will not rain.  It is a reasonable question with a well-defined answer: The sun will shine with probability $x$, and it will snow with probability $1-x$.  But this does not mean that it cannot rain.  If the end of time is a real possibility, then it cannot be prevented just by refusing to ask about it.

\item{Q: In some measures, the cutoff is taken to later and later times.  In this limit, the probability to encounter the end of time surely approaches 0?}

A: No.  In all known measures of this type, an attractor regime is reached where the number of all types of events grows at the same exponential rate, including observers who see 1 o'clock.  The fraction of these observers who also see 2 o'clock before the cutoff approaches a constant less than unity, as will be shown in Sec.~\ref{sec-limit}.

\item Q: But as the cutoff is taken to later times, any given observer's entire history is eventually included. Isn't this a contradiction?

A: No. We do not know which observer we are, so we cannot identify with any particular observer. (If we could, there would be no need for a measure.) Rather, we consider all observers in a specified class, and we define probabilities in terms of the relative frequency of different observations made by these observers.

\item Q: If I looked at what happened on Earth up to the present time (my ``cutoff''), I would find not only records of past clocks that struck both 1 and 2, but also some recently manufactured clocks that have struck 1 but not yet 2.  I could declare that the latter represent a new class of clocks, namely clocks whose existence is terminated by my ``cutoff''.  But I know that this class is fake: it wasn't there before I imposed the ``cutoff''.  Surely, the end of time in eternal inflation is also an artifact that ought to be ignored?

A: Only a finite number of clocks will ever be manufactured on Earth.  Probabilities are given not in terms of the sample picked out by your ``cutoff'', but by relative frequencies in the entire ensemble.  If every clock ever built (in the past or future) strikes both 1 and 2, then the probability for a randomly chosen clock to undergo a different history vanishes, so we may say confidently that the ``cutoff'' has introduced an artifact.  In eternal inflation, however, the cutoff cannot be removed.  Otherwise, we would revert to a divergent multiverse in which relative frequencies are not well-defined.  The cutoff defines not a sample of a pre-existing ensemble; it defines the ensemble.  This is further discussed in Sec.~\ref{sec-artifact}.

\item Q: Why not modify the cutoff to include 2 o'clock?

A: This is a possibility.  If we deform the cutoff hypersurface so that it passes through no matter system, then nothing will run into the end of time.  It is not clear whether this type of cutoff can be obtained from any well-defined prescription.  At a minimum, such a prescription would have to reference the matter content of the universe explicitly in order to avoid cutting through the world volumes of matter systems.  In this paper, we consider only cutoffs defined by a purely geometric rule, which take no direct account of matter.  

\end{itemize}

In Sec.~\ref{sec-GV}, we discuss an apparent paradox that is resolved by the nonzero probability for time to end.

Any conclusion is only as strong as the assumptions it rests on.  The reader who feels certain that time cannot end may infer that at least one of the following assumptions are wrong: (1) the universe is eternally inflating; (2) we may restrict attention to a finite subset of the eternally inflating spacetime, defined by a purely geometric prescription; and (3) probabilities are computed as relative frequencies of outcomes in this subset, Eq.~(\ref{eq-p1p2}).   We discuss these assumptions in Sec.~\ref{sec-assumptions}.

In Sec.~\ref{sec-observation}, we discuss whether, and how, the
nonzero probability for the end of time may be observed.  We point out
that known predictions of various measures actually arise from the
possibility that time can end.  On the problematic side, this includes
the famous youngness paradox of the proper time cutoff; on the
successful side, the prediction of the cosmological constant from the
causal patch cutoff. 

In Sec.~\ref{sec-interpretation}, we discuss how the end of time fits
in with the rest of physics.  This depends on the choice of cutoff.
With the causal patch cutoff, there may be a relatively palatable
interpretation of the end of time which connects with the ideas of
black hole complementarity.  The boundary of the causal patch is a
kind of horizon, which can be treated as an object with physical
attributes, including temperature.  Matter systems that encounter the
end of time are thermalized at this horizon.  This is similar to an
outside observer's description of a matter system falling into a black
hole. What is radically new, however, is the statement that $we$ might
experience thermalization upon crossing the black hole horizon.

This work was inspired by discussions with Alan Guth, who first described to us the paradox mentioned in section \ref{sec-GV}.  We understand that Guth and Vanchurin will be publishing their own conclusions~\cite{GutVanTA}.   In taking seriously the incompleteness of spacetime implied by geometric cutoffs, our conclusion resembles a viewpoint suggested earlier by Ken Olum~\cite{OluPC}.

\section{The probability for time to end}
\label{sec-rate}

The phenomenon that time can end is universal to all geometric cutoffs.  But the rate at which this is likely to happen, per unit proper time $\tau$ along the observer's worldline, is cutoff-specific.  We will give results for five measures.   

\paragraph{Causal patch} The causal patch cutoff~\cite{Bou06} restricts attention to the causal past of the endpoint of a single worldline (see Fig.~\ref{fig-end}). Expectation values are computed by averaging over initial conditions and decoherent histories in the causal patch. The end of time, in this case, is encountered by systems that originate inside the causal patch but eventually exit from it.  

Our universe can be approximated as a flat FRW universe with metric
\begin{equation}
ds^2= - d\tau^2 +a(\tau)^2(d\chi^2 + \chi^2 d\Omega^2)~.
\end{equation}
Observers are approximately comoving ($d\chi/d\tau=0$).   We assume that the decay rate of our vacuum, per unit four-volume, is much less than $t_\Lambda^{-4}$.  Then the decay can be neglected entirely in computing where the boundary of the causal patch intersects the equal time surfaces containing observers.  The boundary is given by the de~Sitter event horizon:
\begin{equation}
\chi_{E}(\tau) = \int_{\tau}^{\infty}{\frac{d\tau'}{a(\tau')}}~.
\label{eq-chieh}
\end{equation}

We consider all observers currently within the horizon: $\chi<\chi_{E}(\tau_0)$, with $\tau_0= 13.7$ Gyr.  This corresponds to a comoving volume $V_{\rm com}=(4\pi/3) \chi_{E}(\tau_0)^3$.  Observers located at $\chi$ leave the patch at a time $\tau'$ determined by inverting Eq.~(\ref{eq-chieh}); in other words, they reach the end of time at $\Delta\tau \equiv \tau'-\tau_0$ from now.  An (unnormalized) probability distribution over $\Delta \tau$ is obtained by computing the number of observers that leave the causal patch at the time $\tau_0+\Delta\tau$:
\begin{equation}
\frac{dp}{d\Delta\tau}\propto \frac{4\pi \chi_E(\tau_0+\Delta\tau)^2}{a(\tau_0+\Delta\tau)}~.
\label{eq-d}
\end{equation}
We compute $a(\tau)$ numerically using the best-fit cosmological parameters from the WMAP5 data combined with SN and BAO~\cite{WMAP5}.  From the distribution (\ref{eq-d}), we may obtain both the median and the expectation value for $\Delta\tau$.  We find that the expected amount of proper time left before time ends is
\begin{equation}
\langle\Delta\tau\rangle=5.3\,\mbox{Gyr}~.
\end{equation}
Time is unlikely to end in our lifetime, but there is a 50\% chance that time will end within the next 3.7 billion years.

\paragraph{Light-cone time}  

The light-cone time of an event is defined in terms of the volume of its future light-cone on the future boundary of spacetime~\cite{Bou09,BouYan09,BouFre10}.  The light-cone time cutoff requires that we only consider events prior to some light-cone time $t_0$; then the limit $t_0\to\infty$ is taken.  It can be shown that the light-cone time cutoff is equivalent to the causal patch cutoff with particular initial conditions~\cite{BouYan09}. Thus, the probability for an observer to encounter the end of time is the same as for the causal patch cutoff.

\paragraph{Fat geodesic}
The fat geodesic cutoff considers a fixed proper volume $4\pi d^3/3$ near a timelike geodesic~\cite{BouFre08b}.  To compute probabilities, one averages over an ensemble of geodesics orthogonal to an initial hypersurface whose details will not matter.  One can show that the geodesics quickly become comoving after entering a bubble of new vacuum.  Since our vacuum is homogeneous, we may pick without loss of generality a fat geodesic at $\chi =0$.  We shall neglect the effects of local gravitational collapse and approximate the universe as expanding homogeneously.  Equivalently, we take the proper distance $d$ to be small compared to the present curvature scale of the universe but large compared to the scale of clusters.  These approximations are not essential, but they will simplify our calculation and save us work when we later consider the scale factor cutoff.

We should only consider observers who are currently ($\tau_0=13.7$ Gyr) within the fat geodesic, with $\chi<d/ a(\tau_0) $.  An observer leaves the geodesic a time $\Delta \tau$ later, with $\chi=d/a(\tau_0+\Delta\tau)$.  The unnormalized probability distribution over $\Delta \tau$ is 
\begin{equation}
\frac{dp}{d\Delta\tau}\propto 4\pi  \frac{d^3 (da/d\tau)_{\tau_0+\Delta\tau}}{a(\tau_0+\Delta\tau)^4}~.
\label{eq-e}
\end{equation}
From this distribution, we find numerically that the expected amount of proper time left before the end of time is $5$ Gyr. There is a 50\% chance that time will end within the next 3.3 billion years.

While the result is similar, there is an important formal difference
between the fat geodesic and causal patch cutoffs.  The boundary of
the fat geodesic is a timelike hypersurface, from which signals can
propagate into the cutoff region.  Boundary conditions must therefore
be imposed. When a system leaves the fat geodesic, time ends from its
own point of view.  But an observer who remains within the cutoff
region continues to see the system and to communicate with it.  The
image of the system and its response to any communications are encoded
in data specified on the timelike boundary.  In practice, the simplest
way to determine these boundary conditions is to consider the global
spacetime and select a fat geodesic from it.  This means that the fat
geodesic is not a self-contained description.  The content of the
causal patch, by contrast, can be computed from its own initial
conditions without reference to a larger spacetime region.

\paragraph{Scale factor time} Scale factor time is defined using a congruence of timelike geodesics orthogonal to some initial hypersurface in the multiverse: $dt\equiv H d\tau$, where $\tau$ is the proper time along each geodesic and $3H$ is the local expansion of the congruence.  This definition breaks down in nonexpanding regions such as dark matter halos; attempts to overcome this limitation (e.g., Ref.~\cite{DGSV08}) remain somewhat ad-hoc. Here we use for $H$ the Hubble rate of a completely homogeneous universe whose density agrees with the average density of our universe. This does not yield a precise and general cutoff prescription, but it allows us to compute an approximate rate at which we are likely to encounter the cutoff:  in an everywhere-expanding timelike geodesic congruence, the scale factor time cutoff is equivalent to the fat geodesic cutoff ~\cite{BouFre08b}. Hence it gives the same rate for time to end as the fat geodesic cutoff.

\paragraph{Proper time}
In the proper time cutoff, the characteristic timescale is the shortest Hubble time of all eternally inflating vacua.  In a realistic landscape, this is microscopically short, perhaps of order the Planck time ~\cite{BouFre07}.  Thus, time would be overwhelmingly likely to end in the next second:
\begin{equation}
{dp \over d \Delta\tau} \approx t_{\rm Pl}^{-1}~.
\label{eq-pt}
\end{equation}
This is the famous ``youngness paradox'' in a new guise.  The cutoff predicts that our observations have superexponentially small probability, and that most observers are ``Boltzmann babies'' who arise from quantum fluctuations in the early universe.  Thus, this measure is already ruled out phenomenologically at a high level of confidence~\cite{LinLin96,Gut00a,Gut00b,Gut04,Teg05,Lin07,Gut07,BouFre07}.

\section{Objections}
\label{sec-ob}
Our intuition rebels against the conclusion that spacetime could simply cease to exist.  In the introduction, we answered several objections that could be raised against the end of time.  In this section, we will discuss two of these arguments in more detail.

\subsection{Time cannot end in a late-time limit}
\label{sec-limit}

In some measure proposals, such as the proper time cutoff~\cite{Lin86a,Lin06}, the scale factor time cutoff~\cite{DGSV08}, and the light-cone time cutoff~\cite{Bou09}, a limit is taken in which the cutoff region is made larger and larger as a function of a time parameter $t_0$: 
\begin{equation}
p_1/p_2=\lim_{t_0\to\infty} N_1(t_0) / N_2(t_0)~.  
\end{equation}
Naively one might expect the cutoff observers to be an arbitrarily small fraction of all observers in the limit $t_0 \to \infty$. This turns out not to be the case. 

One finds that the number of events of type $I$ that have occurred prior to the time $t$ is of the form
\begin{equation}
N_I(t)=\check N_I \exp(\gamma t)+O(\sigma t)~,
\label{timedep}
\end{equation}
with $\sigma<\gamma\approx 3$.  Thus, the growth approaches a universal exponential behavior at late times~\cite{GarSch05,BouFre07,BouYan09}, independently of initial conditions. The ratio $N_1/N_2$ appearing in Eq.~(\ref{eq-p1p2}) remains well-defined in the limit $t_0\to\infty$, and one obtains
\begin{equation}
\frac{p_1}{p_2}=\frac{\check N_1}{\check N_2}~.
\end{equation}
The constants $\check N_I$, and thus the relative probabilities, depend on how the time variable is defined; we will discuss some specific choices below.

Suppose that observers live for a fixed global time interval $\Delta t$. Then a person dies before time $t$ if and only if he was born before $t - \Delta t$.  Therefore the number of births $N_b$ is related to the number of deaths $N_d$ by
\begin{equation}
N_d (t) = N_b (t - \Delta t)
\end{equation}
Using the time dependence of $N_b$ given in \eqref{timedep}, this can be rewritten
\begin{equation}
\frac{N_d(t)}{N_b(t)}\approx \exp(- \gamma \Delta t)~,
\end{equation}
up to a correction of order $e^{(\sigma-\gamma)t}$ which becomes negligible in the late time limit.  Thus, the fraction of deaths to births does not approach unity as the cutoff is taken to infinity.  The fraction of observers whose lives are interrupted by the cutoff is
\begin{equation}
{N_c \over N_b} = 1 - \exp(-\gamma \Delta t)
\label{obsfrac}
\end{equation}
where $N_c=N_b-N_d$ is the number of cutoff observers.

Since \eqref{obsfrac} is true for any time interval $\Delta t$, it is
equivalent to the following simple statement: any system has a
constant probability to encounter the end of time given by
\begin{equation}
{dp \over d t} = \gamma \approx 3
\label{probdist}
\end{equation}
This result can be interpreted as follows.  Due to the steady state behavior of eternal inflation at late times, there is no way to tell what time it is. The exponential growth \eqref{timedep} determines a $t_0$-independent probability distribution for how far we are from the cutoff, given by \eqref{probdist}.

\subsection{The end of time is an artifact}
\label{sec-artifact}

Could it be that observers who run into the cutoff are an artifact, not to be taken seriously as a real possibility?  Certainly they would not exist but for the cutoff.  Yet, we argue that cutoff observers are a real possibility, because {\em there is no well-defined probability distribution without the cutoff; in particular, only the cutoff defines the set of allowed events}.   In order to convince ourselves of this, it is instructive to contrast this situation with one where a cutoff may introduce artifacts.  We will consider two finite ensembles of observers, without reference to eternal inflation.   We then restrict attention to a portion of each ensemble, defined by a cutoff.  We find that this sample looks the same in both ensembles, and that it contains observers that run into the cutoff.  In the first ensemble, these observers are an artifact of sampling; in the second, they are real.  We will then explain why eternal inflation with a cutoff is different from both examples.

\paragraph{A cutoff on a finite ensemble defines a sample}

Consider a civilization which begins at the time $t=0$ and grows exponentially for a finite amount of time.  (We will make no reference to a multiverse in this example.)  Every person is born with a watch showing 1 o'clock at the time of their birth, when they first look at it.  One hour later, when they look again, the watch shows 2 o'clock; immediately thereafter the person dies.  After the time $t_*\gg 1$ hour, no more people are born, but whoever was born before $t_*$ gets to live out their life and observe 2 o'clock on their watch before they die.  In this example, there is a well-defined, finite ensemble, consisting of all observers throughout history and their observations of 1 and 2.  The ensemble contains an equal number of $1$'s and $2$'s.  Every observer in the ensemble sees both a $1$ and a $2$, each with 100\% probability.  No observer meets a premature demise before seeing $2$. 

Now suppose that we do not know the full ensemble described above.  Instead, we are allowed access only to a finite sample drawn from it, namely everything that happened by the time $t$, with 1 hour $\ll t<t_*$.  This sample contains many observers who died before $t$; each of them will have seen both 1 and 2. We refer to these as ``histories" of type $12$.  It also contains observers (those who were born after $t-1$ hour) who are still alive.  Each of them has seen 1 but not yet 2, by the time $t$, which we refer to as a history of type $1-$.  What do we say about these observers?  Should we declare that there is a nonzero probability that an observer who sees 1 does not live to see 2?   In fact, a finite sample of a larger, unknown ensemble allows us to draw no conclusion of this kind, because we have no guarantee that our sampling of the ensemble is fair.  The true set of outcomes, and their relative frequency, is determined only by the full ensemble, not by our (possibly unfaithful) sample of it.  Similarly, if we had a considered a more complicated system, such as observers with watches of different colors, etc., the relative frequency of outcomes in any subset need not be the same as the relative frequency in the full ensemble, unless we make further assumptions.

If we examined the full ensemble, we could easily verify that every observer who sees 1 also lives to see 2.  Thus we would learn that $1-$ was an artifact of our sampling: imposing a cutoff at fixed time produced a new class of events that does not exist (or more precisely, whose relative frequency drops to zero) once we take the cutoff away.  Armed with this knowledge, we could then invent an improved sampling method, in which the $1-$ cases are either excluded, or treated as $12$ events.

As our second example, let us consider a civilization much like the previous one, except that it perishes not by a sudden lack of births, but by a comet that kills everyone at the time $t_*$.  This, too, gives rise to a finite, well-defined ensemble of observations.  But unlike in the previous example, there is a larger number of $1$'s than $2$'s: not every observer who sees a $1$ lives to see a $2$.  Thus, the probabilities for the histories $12$ and $1-$ satisfy $p_{1-}>0$, $p_{12}<1$.  Indeed, if we choose parameters so the population grows exponentially on a timescale much faster than 1 hour, most people in history who see $1$ end up being killed by the comet rather than expiring naturally right after seeing $2$; that is, $p_{12}=1-p_{1-}\ll 1$ in this limit.   

Again, we can contemplate sampling this ensemble, i.e., selecting a subset, by considering everything that happened prior to the time $t<t_*$.  Note that this sample will look identical to the finite-time sample we were given in the previous example.  Again, we find that there are apparently events of type $1-$, corresponding to observers who have seen 1 but not 2 by the time $t$.  But in this example, it so happens that (i) events of type $1-$ actually do exist in the full ensemble, i.e., have nonzero true relative frequency; and (ii) assuming exact exponential growth, our sample is faithful: the relative frequency of $1-$ vs.\ $12$ in the sample (observers prior to $t$) is the same as in the full ensemble (observers in all history, up to $t_*$).\footnote{Actually it is faithful only in the limit as $t$ is much greater than the characteristic growth timescale of the civilization, because of the absence of any observers prior to $t=0$.}

We learn from the above two examples that a subset of an ensemble need not yield reliable quantitative information about the relative frequencies of different outcomes, or even qualitative information about what the allowed outcomes are.  All of this information is controlled only by the full ensemble.  In both examples, the set of events that occured before the time $t<t_*$ contain events of type $1-$.  But in the first example, these events are a sampling artifact and their true probability is actually 0.  In the second example, $1-$ corresponds to a real possibility with non-zero probability.

\paragraph{The cutoff in eternal inflation defines the ensemble}

Now let us return to eternal inflation.  In order to regulate its divergences, we define a cutoff that picks out a finite spacetime region, for example the region prior to some constant light-cone time $t$.  Naively, this seems rather similar to the examples above, where we sampled a large ensemble by considering only the events that happened prior to a time $t<t_*$.  But we learned that such samples cannot answer the question of whether the histories of type $1-$ are real or artifacts.  To answer this question, we had to look at the full ensemble.  We found in the first example that $1-$ was real, and in the second that $1-$ was an artifact, even though the sample looked exactly the same in both cases.  In eternal inflation, therefore, we would apparently need to ``go beyond the cutoff'' and consider the ``entire ensemble'' of outcomes, in order to decide whether $1-$ is something that can really happen.

But this is impossible: the whole point of the cutoff was to {\em define\/} an ensemble.  An infinite set is not a well-defined ensemble, so the set we obtained by imposing a cutoff {\em is\/} the most fundamental definition of an ensemble available to us.  We can argue about which cutoff is correct: light-cone time, scale factor time, the causal patch, etc.  But whatever the correct cutoff is, its role is to define the ensemble.  It cannot be said to select a sample from a larger ensemble, namely from the whole multiverse, because this larger ensemble is infinite, so relative abundances of events are not well-defined.  If they were, we would have had no need for a cutoff in the first place.

\section{The Guth-Vanchurin paradox}
\label{sec-GV}

Another way to see that the end of time is a real possibility is by verifying that it resolves a paradox exhibited by Guth and Vanchurin~\cite{GutVanPC}. Suppose that before you go to sleep someone flips a fair coin and, depending on the result, sets an alarm clock to awaken you after either a short time, $\Delta t \ll 1$, or a long time $\Delta t \gg 1$.  Local physics dictates that there is a 50\% probability to sleep for a short time since the coin is fair. Now suppose you have just woken up and have no information about how long you slept. It is natural to consider yourself a typical person waking up. But if we look at everyone who wakes up before the cutoff, we find that there are far more people who wake up after a short nap than a long one. Therefore, upon waking, it seems that there is no longer a 50\% probability to have slept for a short time.

How can the probabilities have changed?  If you accept that the end of time is a real event that could happen to you, the change in odds is not surprising: although the coin is fair, some people who are put to sleep for a long time never wake up because they run into the end of time first. So upon waking up and discovering that the world has not ended, it is more likely that you have slept for a short time. You have obtained additional information upon waking---the information that time has not stopped---and that changes the probabilities.

However, if you refuse to believe that time can end, there is a contradiction.  The odds cannot change unless you obtain additional information.  But if all sleepers wake, then the fact that you woke up does not supply you with new information.

Another way to say it is that there are two reference classes one could consider. When going to sleep we could consider all people falling asleep; 50\% of these people have alarm clocks set to wake them up after a short time. Upon waking we could consider the class of all people waking up; most of these people slept for a short time. These reference classes can only be inequivalent if some members of one class are not part of the other. This is the case if one admits that some people who fall asleep never wake up, but not if one insists that time cannot end.

\section{Discussion}
\label{sec-discussion}

Mathematically, the end of time is the statement that our spacetime manifold is extendible, i.e., that it is isometric to a proper subset of another spacetime. Usually, it is assumed that spacetime is inextendable~\cite{Wald}.  But the cutoffs we considered regulate eternal inflation by restricting to a subset of the full spacetime.  Probabilities are fundamentally defined in terms of the relative abundance of events and histories in the subset.  Then the fact that spacetime is extendible is itself a physical feature that can become part of an observer's history.  Time can end.

\subsection{Assumptions}
\label{sec-assumptions}

We do not know whether our conclusion is empirically correct.  What we have shown is that it follows logically from a certain set of assumptions.  If we reject the conclusion, then we must reject at least one of the following propositions:

\paragraph{Probabilities in a finite universe are given by relative frequencies of events or histories} 

This proposition is sometimes called the assumption of typicality.  It forces us to assign a nonzero probability to encountering the end of time if a nonzero fraction of observers encounter it. 

Even in a finite universe one needs a rule for assigning relative probabilities to observations.  This need is obvious if we wish to make predictions for cosmological observations.  But a laboratory experiment is a kind of observation, too, albeit one in which the observer controls the boundary conditions.  A comprehensive rule for assigning probabilities cannot help but make predictions for laboratory experiments in particular.  However, we already have a rule for assigning probabilities in this case, namely quantum mechanics and the Born rule, applied to the local initial conditions prepared in the laboratory.  This must be reproduced as a special case by any rule that assigns probabilities to all observations~\cite{BouFre07}.  A simple way to achieve this is by defining probabilities as  ratios of the expected number of instances of each outcome in the universe, as we have done in Eq.~(\ref{eq-p1p2}).

\paragraph{Probabilities in an infinite universe are defined by a
  geometric cutoff} This proposition states that the infinite
spacetime of eternal inflation must be rendered finite so that the
above frequency prescription can be used to define probabilities.
Moreover, it states that a finite spacetime should be obtained by
restricting attention to a finite subset of the infinite
multiverse.\footnote{We have considered measures in which the cutoff
  is completely sharp, i.e., described by a hypersurface that divides
  the spacetime into a region we keep and a region we discard.  In
  fact this is not essential.  One could smear out the cutoff by
  assigning to each spacetime event a weight that varies smoothly from
  1 to 0 over some region near the cutoff surface.  There would still
  be a finite probability for time to end.}  It is possible that the
correct measure cannot be expressed in a geometric form.  Imagine, for
instance, a measure that makes ``exceptions'' for matter systems that
come into existence before the cutoff, allowing all events in their
world volume to be counted.  A purely geometric prescription would
have chopped part of the history off, but in this measure, the cutoff
surface would be deformed to contain the entire history of the system.
Such a cutoff would depend not only on the geometry, but also on the
matter content of spacetime.\footnote{We have not attempted to prove
  this statement, so it should be considered an additional assumption.
  Because the metric has information about the matter content, we
  cannot rule out that a geometric measure could be formulated whose
  cutoff surfaces never intersect with matter. It seems unlikely to us
  that such a cutoff could select a finite subset of the
  multiverse. A related possibility would be to define a global time
  cutoff such that typical observers live farther and farther from the
  cutoff in the limit as $t\to\infty$.  This would invalidate our
  analysis in Sec.~\ref{sec-limit}, which assumed exponential growth
  in $t$.}  A more radical possibility is that the measure may not
involve any kind of restriction to a finite portion of spacetime.  For
example, Noorbala and Vanchurin~\cite{NooVan10}, who exhibit a paradox
similar to that described in Sec.~\ref{sec-GV}, but do not allow for
the possibility that time can end, advocate a nongeometric type of
measure. If such a prescription could be made well-defined and
consistent with observation (which seems unlikely to us), then one
might escape the conclusion that time can end. Similarly, Winitzki~\cite{Win08a,Win08b,Win08c} defined a measure where only finite spacetimes are
considered, and in this measure there is no novel catastrophe like the
end of time.

\paragraph{The universe is eternally inflating}
To prove this proposition wrong would be a dramatic result, since it would seem to require a kind of fundamental principle dictating that Nature abhors eternal inflation.  After all, eternal inflation is a straightforward consequence of general relativity, assuming there exists at least one sufficiently long-lived field theory vacuum with positive vacuum energy (a de~Sitter vacuum).  This assumption, in turn, seems innocuous and is well-motivated by observation: (1) The recent discovery of accelerated expansion~\cite{Rie98,Per98}, combined with the conspicuous lack of evidence that dark energy is not a cosmological constant~\cite{WMAP5}, suggests that our own vacuum is de~Sitter.  If this is the case, the universe must be eternally inflating.  (2) Slow-roll inflation solves the horizon and flatness problems.  Its generic predictions agree well with the observed CMB power spectrum.  But slow-roll inflation requires a sufficiently flat scalar field potential.  Eternal inflation requires only a local minimum and so is less fine-tuned.  How could we consider slow-roll inflation, but exclude eternal inflation? --- There are also theoretical motivations for considering the existence of de~Sitter vacua: (3) In effective field theory, there is nothing special about potentials with a positive local minimum, so it would be surprising if they could not occur in Nature. (4) String theory predicts a very large number of long-lived de~Sitter vacua\cite{BP,KKLT,DenDou04b}, allowing for a solution of the cosmological constant problem and other fine-tuning problems.


\subsection{Observation}
\label{sec-observation}

If we accept that time can end, what observable implications does this
have?  Should we expect to see clocks or other objects suddenly
disappear?  In measures such as scale factor time or light-cone time,
the expected lifetime of stable systems is of order 5 billion years
right now, so it would be very unlikely for the end of time to occur
in, say, the next thousand years.  And even if it did occur, it would
not be observable.  Any observer who would see another system running
into the end of time is by definition located to the causal future of
that system.  If the cutoff surface is everywhere spacelike or null,
as is the case for the light-cone time cutoff and the causal patch
cutoff, then the observer will necessarily run into the cutoff before
observing the demise of any other system.

Though the end of time would not be observable, the fact that time has
{\em not\/} ended certainly is observable.  If a theory assigns
extremely small probability to some event, then the observation of
this event rules out the theory at a corresponding level of
confidence.  This applies, in particular, to the case where the event
in question is time-not-having-ended.  For example, Eq.~(\ref{eq-pt})
shows that the proper time measure is thus falsified. 

An observation which indirectly probes the end of time is the value of the cosmological constant.  For definiteness consider the causal patch measure, which predicts a coincidence between the time when the observers live and the time when the cosmological constant beings to dominate the expansion of the universe, $t_\Lambda \sim t_{\rm obs}$.  This represents the most important phenomenological success of the measure, and we will now argue that it is tied intimately to the end of time.

The most likely value of the cosmological constant is the one which leads to the most observers inside the causal patch.  We will assume that there are a constant number of observers per unit mass, and will imagine scanning the possible values of $t_\Lambda \sim 1/\sqrt{\Lambda}$ with $t_{\rm obs}$ held fixed.  It's most useful to think of the distribution of values of $\log t_\Lambda$, where the preferred value is largely determined by two competing pressures.  First, since the prior probability is flat in $\Lambda$, there is an exponential pressure in $\log t_\Lambda$ toward lesser values.  Second, if $t_\Lambda < t_{\rm obs}$ there is an exponential pressure in $t_\Lambda$ (superexponential in $\log t_\Lambda$) toward greater values.  This is a simple consequence of the fact that all matter is expelled from the causal patch at an exponential rate after vacuum domination begins.  These two pressures lead to $t_{\rm obs} \sim t_\Lambda$.

The end of time is implicitly present in this argument.  Suppose there are two generations of observers, one living at $t_\Lambda$ and another at $10t_\Lambda$.  Even if local physics says that there are the same number of observers per unit mass in each generation, the second generation must be atypical, and hence have fewer members, if the prediction for the cosmological constant is to remain valid.  Where are the missing members of the second generation?  The answer is that time has ended for them.  They are not counted for the purposes of any calculation, and so they do not exist.  Clearly, the setup is identical to the observers who see 1 o'clock and 2 o'clock discussed above.

In this paper, we have considered sharp geometric cutoffs. However, intuition from AdS/CFT~\cite{GarVil08, Bou09, BouFre10} suggests that the cutoff should not be a
completely sharp surface, but should be smeared out over a time of
order $t_\Lambda$. If the cutoff is smeared, there could be observable
consequences of approaching the end of time; the details would depend
on the precise prescription for smearing the cutoff.

\subsection{Interpretation}
\label{sec-interpretation}

The notion that time can come to an end is not completely new.  Space and time break down at at singularities, which are guaranteed to arise in gravitational collapse~\cite{HawEll}.   But our conclusion is more radical: the world can come to an end in any spacetime region, including regions with low density and curvature, because spacetime is incomplete.  

One might speculate that semiclassical gravity breaks down on very large time scales, say $t_\Lambda^3$, the evaporation time for a large black hole in de~Sitter space, or $\exp(\pi t_\Lambda^2)$, the recurrence time in de~Sitter space.  But in the most popular measures, we are likely to encounter the end of time on the much shorter timescale $t_\Lambda$.  Perhaps one could invent a new cutoff that would push the end of time further into the future.  But there is no well-motivated candidate we are aware of, and, as we have discussed, one would be likely to lose some of the phenomenological success of the measures in solving, e.g.,  the cosmological constant problem.

How can we make sense of our conclusion?  Is there a way of thinking about it that would make us feel more comfortable about the end of time?  Does it fit in with something we already know, or is this a completely new phenomenon?  The answer to this question turns out to depend somewhat on which cutoff is used.

\paragraph{All measures}  One way to interpret the end of time is to
imagine a computer just powerful enough to simulate
the cutoff portion of the eternally inflating spacetime.  The
simulation simply stops at the cutoff.  If the measure involves taking
a late time limit, then one can imagine building larger and larger
computers that can simulate the spacetime until a later cutoff.  These
computers can be thought of as the definition of the cutoff theory,
much in the same way that lattice gauge theory is used. There is no
physical significance to any time after the cutoff.\footnote{Ken Olum has pointed out for some time that one way to interpret a geometric cutoff is that ``we are being simulated by an advanced civilization with a large but finite amount of resources, and at some point the simulation will stop.''  The above interpretation adopts this viewpoint (minus the advanced civilization).}  This is an interesting rephrasing of the statement of the end of time, but it does not seem to mitigate its radicality.

\paragraph{Causal patch only} Our result appears to admit an
intriguing interpretation if the causal patch measure is used.  The
original motivation for the causal patch came from black hole
complementarity~\cite{SusTho93}.  Consider the formation and
evaporation of a black hole in asymptotically flat space.  If this
process is unitary, then the quantum state describing the collapsing
star inside the black hole is identical to the state of the Hawking
radiation cloud.  Since these two states are spacelike separated, two
copies of the quantum state exist at the same time.  But before the
star collapsed, there was only one copy.  This amounts to ``quantum
xeroxing'', which is easily seen to conflict with quantum mechanics.

A way around this paradox is to note there is no spacetime point whose past light-cone contains both copies.  This means that no experiment consistent with causality can actually verify that xeroxing has taken place.  Thus, the paradox can be regarded as an artifact of a global viewpoint that has no operational basis.  A theory should be capable of describing all observations, but it need not describe more than that.  Geometrically, this means that it need not describe any system that cannot fit within a causal patch.  What the xeroxing paradox teaches us is that we {\em must\/} not describe anything larger than the causal patch if we wish to avoid inconsistencies in spacetimes with black holes.  

But once we reject the global description of spacetime, we must reject it whether or not black holes are present.  In many cosmological solutions, including eternal inflation, the global spacetime is not accessible to any single experiment.  This motivated the use of the causal patch as a cutoff to regulate the infinities of eternal inflation~\cite{BouFre06a,Bou06}.  

Let us return to the black hole spacetime and consider the causal patch of an outside observer. This patch includes all of the spacetime except for the interior of the black hole.  As Susskind has emphasized, to an outside observer, the causal patch is a consistent representation of the entire world.  The patch has a boundary, the stretched horizon of the black hole.  This boundary behaves like a physical membrane, endowed with properties such as temperature, tension, and conductivity.  When another observer falls into the black hole, the outside observer would say that he has been thermalized at the horizon and absorbed into the membrane degrees of freedom.  Later the membrane evaporates and shrinks away, leaving behind a cloud of radiation.   

It is very important to understand that this really is the unique and complete description of the process from the outside point of view; the black hole interior does not come into it.  The process is no different, in principle, from throwing the second observer into a fire and watching the smoke come out.  Any object is destroyed upon reaching the horizon.  Yet, assuming that the black hole is large, the infalling observer would not notice anything special when crossing the horizon.  There is no contradiction between these two descriptions, since they agree as long as the two observers remain in causal contact.  Once they differ, it is too late for either observer to send a signal to the other and tell a conflicting story.

The end of time in the causal patch is an effect that fits well with
the outside observer's description.  When the infalling observer
enters the black hole, he is leaving the causal patch of the outside
observer.  In the language of the present paper, the outside observer
defines a particular causal patch, and the inside observer encounters
the end of time when he hits the boundary of this patch.  We now see
that there is a different, more satisfying interpretation: the inside
observer is thermalized at the horizon.  This interpretation invokes a
relatively conventional physical process to explain why the inside
observer ceases to exist.  Time does not stop, but rather, the
observer is thermalized.  His degrees of freedom are merged with those
already existing at the boundary of the causal patch, the horizon.

If this interpretation is correct, it can be applied to black holes that form in the eternally inflating universe, where it modifies the theory of the infalling observer.  It is no longer certain that an infalling observer will actually make it to the horizon, and into the black hole, to perish only once he hits the future singularity.  Instead, time might end before he enters the black hole.   How is this possible?

In the traditional discussion of black hole complementarity, one picks an observer and constructs the associated causal patch.   It is impossible, by construction, for an observer to leave his own patch.  In other words, time cannot end if we live in a causal patch centered on our own worldline.  In eternal inflation, however, one first picks a causal patch; then one looks for observers in it.  Some of these observers will be closer to the boundary and leave the patch sooner than others, who happen to stay in the patch longer.  Equivalently, suppose we do want to begin by considering observers of a given type, such as an observer falling towards a black hole.  To compute probabilities, we must average over all causal patches that contain such an observer.  In some patches the observer will be initially far from the boundary, in others he will hit the boundary very soon.  This yields a probability distribution for the rate at which time ends.

Suppose, for example, that we attempted to jump into a black hole of
mass $m$ in our own galaxy (and neglect effects of gravitational tidal
forces, matter near the black hole, etc.). Using the ensemble of
causal patches defined in Ref.~\cite{BouYan09}, one finds that time would probably end
before we reach the horizon, with probability $1-O(m /
t_\Lambda)$.  This probability is overwhelming if the black hole is much smaller than the cosmological horizon.

\acknowledgments We would like to particularly thank Adam Brown and
Alan Guth for very influential discussions. We also thank David Berenstein, Steve
Shenker, Lenny Susskind, and Vitaly
Vanchurin for helpful discussions.  This work was supported by the Berkeley
Center for Theoretical Physics, by the National Science Foundation
(award number 0855653), by the Institute for the Physics and
Mathematics of the Universe, by fqxi grant RFP2-08-06, and by the US
Department of Energy under Contract DE-AC02-05CH11231. VR is supported
by an NSF graduate fellowship.

\bibliographystyle{utcaps}
\bibliography{all}
\end{document}